\documentclass[journal=jpclcd,manuscript=article]{achemso}
\usepackage{chemformula} % Formula subscripts using \ch{}
\usepackage{amsmath,amssymb,amsfonts,latexsym,graphicx}
\usepackage{geometry}
\usepackage{indentfirst}
\usepackage{fancyhdr}
\usepackage{titletoc}
\usepackage{titlesec}
\usepackage{booktabs}
\usepackage{threeparttable}
\usepackage{multirow}
\usepackage{color}
\usepackage{rotating}
\usepackage[version=4]{mhchem} % Formula subscripts using \ce{}
\setkeys{acs}{maxauthors = 0} % references of many authors

\newcommand{\osinglet}{|{}^1\Psi_{\sigma_{1s_\mathrm{O}}\rightarrow \pi^*_x}\rangle}
\newcommand{\csinglet}{|{}^1\Psi_{\sigma_{1s_\mathrm{C}}\rightarrow \pi^*_x}\rangle}

\usepackage{etoolbox} % needed for patch
\makeatletter
\patchcmd{\acs@contact@details}{E}{*\,E}{}{}
\makeatother

\makeatletter
\def\acs@author@fnsymbol#1{}
\makeatother

\author{Kaixuan Huang$^{2,3,\#}$,
Xiaoxia Cai$^{1,\#}$,
Hao Li$^{3,7,\#}$,
Zi-Yong Ge$^5$,
Ruijuan Hou$^1$,\\
Hekang Li$^3$,
Tong Liu$^{3,4}$,
Yunhao Shi$^{3,4}$,
Chitong Chen$^{3,4}$,
Dongning Zheng$^{3,4,6}$}
\author{Kai~Xu$^{3,4,6,\ast}$}\email{kaixu@iphy.ac.cn}
\author{Zhi-Bo~Liu$^{2,\ast}$}\email{liuzb@nankai.edu.cn}
\author{Zhendong Li$^{1,\ast}$}\email{zhendongli@bnu.edu.cn}
\author{Heng Fan$^{3,4,6,\ast}$}\email{hfan@iphy.ac.cn}
\author{Wei-Hai Fang$^1$}
\affiliation{
\normalsize{$^1${\it Key Laboratory of Theoretical and Computational Photochemistry, Ministry of Education, College of Chemistry, Beijing Normal University, Beijing 100875, China}}\\
\normalsize{$^2${\it The Key Laboratory of Weak Light Nonlinear Photonics, Ministry of Education, Teda Applied Physics Institute and School of Physics, Nankai University, Tianjin 300457, China}}\\
\normalsize{$^3${\it Institute of Physics, Chinese Academy of Sciences, Beijing 100190, China}}\\
\normalsize{$^4${\it School of Physical Sciences, University of Chinese Academy of Sciences, Beijing 100190, China}}\\
\normalsize{$^5${\it Theoretical Quantum Physics Laboratory, RIKEN Cluster for Pioneering Research, Wako-shi, Saitama 351-0198, Japan}}\\
\normalsize{$^6${\it CAS Center for Excellence in Topological Quantum Computation, University of Chinese Academy of Sciences, Beijing 100190, China}}\\
\normalsize{$^7${\it School of Physics, Northwest University, Xi'an 710127, China}}\\
\normalsize{$^{\#}${\it Contributed equally to this work}}\\
}

\title{Variational Quantum Computation of Molecular Linear Response Properties
on a Superconducting Quantum Processor}

\begin{document}

\begin{tocentry}
\includegraphics[width=5cm]{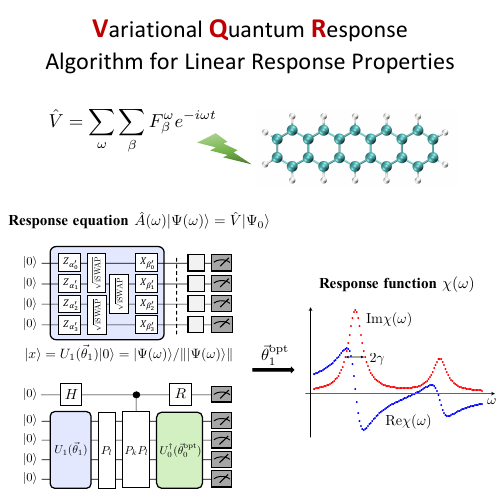}
\end{tocentry}

\begin{abstract}
Simulating response properties of molecules is crucial for interpreting experimental spectroscopies and accelerating materials design. However, it remains a long-standing computational challenge for
electronic structure methods on classical computers.
While quantum computers hold the promise to solve this problem more efficiently in the long run, existing quantum algorithms requiring deep quantum circuits
are infeasible for near-term noisy quantum processors.
Here, we introduce a pragmatic variational quantum response (VQR) algorithm for
response properties, which circumvents the need for deep quantum circuits.
Using this algorithm, we report the first simulation of linear response properties
of molecules including dynamic polarizabilities and absorption spectra on a
superconducting quantum processor.
Our results indicate that a large class of important dynamical properties
such as Green's functions are within the reach of near-term quantum hardware
using this algorithm in combination with
suitable error mitigation techniques.
\end{abstract}

In silico simulation of molecular response properties, such as dynamic polarizabilities,
absorption/emission spectra, and nonlinear optical properties,
is essential in virtual design of new materials with specific properties\cite{marzari2021electronic}.
However, despite decades of effort and tremendous progress in methodological developments,
their accurate and efficient prediction remains
a fundamental challenge for electronic structure methods
on classical computers\cite{helgaker2012recent,norman2018principles}.
On one hand, while density functional theory (DFT) and its time-dependent extension (TD-DFT) have been the workhorse
for simulating structural and response properties of large molecules due to their good efficiency, the limitations of density functional approximations are well-known\cite{cohen2008insights}, especially for strongly correlated systems.
On the other hand, the exact method for solving the many-electron Schr\"{o}dinger equation within a given basis set, namely, full configuration interaction (FCI) or exact diagonalization\cite{helgaker2012recent}, has an exponential scaling with respect
to the system size in both physical memory and computational time.

Quantum computing is promising for simulating interacting many-body systems including molecules\cite{cao2019quantum,mcardle2020quantum,bauer2020quantum,motta2021emerging}.
So far, quantum simulations of molecules on the present noisy intermediate-scale quantum (NISQ)\cite{preskill2018quantum} devices
have almost exclusively focused on the ground state $|\Psi_0\rangle$ and the associated
energy $E_0$\cite{peruzzo2014variational,o2016scalable,kandala2017hardware,hempel2018quantum,sagastizabal2019experimental,arute2020hartree}, which is the very first step towards
simulating molecular properties. Computing (frequency-dependent) linear and nonlinear response properties is far more challenging, because they implicitly involve not only the ground state but also all the excited states\cite{norman2018principles}. While it is possible to compute response properties
using the sum-over-state formula (see Eq. \eqref{eq:SOS}) with individual
excited state computed by existing quantum algorithms\cite{mcclean2017hybrid,colless2018computation,higgott2019variational,nakanishi2019subspace,parrish2019quantum,ollitrault2020quantum}, such approach will quickly become
impractical as the number of excited states grows exponentially with the system size.
Previous quantum algorithms for linear response functions largely rely on computing time-correlation functions using quantum computers\cite{somma2002simulating,chiesa2019quantum,francis2020quantum,sun2021quantum},
which can later be Fourier transformed to the frequency domain on classical computers.
Very recently, they have been realized for spin-$\frac{1}{2}$ Heisenberg models (with up to four sites) on superconducting quantum processors\cite{chiesa2019quantum,francis2020quantum,sun2021quantum}.
However, as implementing the time evolution operator $e^{-i\hat{H}_0t}$
with the molecular Hamiltonian $\hat{H}_0$ requires a formidably large circuit depth,
which at least scales as $O(n^4)$ with respect to the number of qubits $n$\cite{whitfield2011simulation,seeley2012bravyi,hastings2015improving} or $O(n^2)$ with low-rank approximations\cite{motta2021low}
asymptotically, it is difficult to apply these algorithms for molecules on NISQ devices.
Thus, how to compute molecular response properties on near-term quantum hardware remains an open problem.

In this Letter, we describe a pragmatic quantum computational approach
for computing linear response properties, which circumvents the need for
deep quantum circuits. Specifically, we propose efficient quantum circuits
to solve the frequency-domain linear response equation in a variational hybrid quantum-classical way. Together with an error mitigation (EM) strategy based on symmetry projection to mitigate the
impact of noises inherent in NISQ devices, we realize the first
simulation of linear response properties of molecules
including dynamic polarizabilities and absorption spectra
on a programmable superconducting quantum processor\cite{Song2019Generation}.
The present approach can be easily extended to simulate other important
dynamical properties such as Green's functions and nonlinear response properties,
and is also applicable to other platforms such as trapped-ion systems\cite{hempel2018quantum,Nam2020}.

{\it Variational quantum response (VQR) algorithm.}
We consider the computation of the following linear response function, which
characterizes the first order response of molecules to an applied external field
\begin{eqnarray}
\chi(\omega)
=
\sum_{m}\frac{|\langle\Psi_m|\hat{V}|\Psi_0\rangle|^2}
{\omega_{m0}-(\omega+i\gamma)},\label{eq:SOS}
\end{eqnarray}
where $\omega$ represents the frequency of the external field
and the parameter $\gamma$ determines the frequency resolution,
which is usually adjusted to best fit an experimental spectrum of molecules\cite{norman2018principles}.
Depending on the perturbation $\hat{V}$, $\chi(\omega)$ can be
dynamic polarizabilities, magnetic susceptibilities,
Green's functions, etc. It encodes all the information of
the excitation process from the ground state $|\Psi_0\rangle$ to the $m$-th excited state
$|\Psi_m\rangle$ due to the perturbation $\hat{V}$, viz., the excitation energy $\omega_{m0}=E_m-E_0$
and the transition amplitude $|\langle\Psi_m|\hat{V}|\Psi_0\rangle|^2$.
Exact calculations of $\chi(\omega)$ on classical computers are generally intractable, as
the number of excited states scales exponentially with the molecular size\cite{helgaker2012recent}.
To compute $\chi(\omega)$ on quantum computers,
an appropriate fermion-to-qubit mapping\cite{bravyi2002fermionic,seeley2012bravyi}
is first applied to transform $\hat{H}_0$ and $\hat{V}$ for electrons to their counterparts for $n$ qubits,
i.e., $\hat{H}_0=\sum_k h_k P_k$ and $\hat{V}=\sum_k v_k P_k$,
which are linear combinations of Pauli terms $P_k=\sigma^k_0\otimes\sigma^k_1\otimes\cdots\otimes\sigma^k_{n-1}$
with $\sigma_i\in\{I,X,Y,Z\}$. For spin models with a much simpler $\hat{H}_0$,
previous quantum algorithms\cite{somma2002simulating,chiesa2019quantum,francis2020quantum,sun2021quantum}
computed the dynamical correlation function
$C(t)=\langle\Psi_0|\hat{V}^\dagger(t)\hat{V}|\Psi_0\rangle
=\sum_{lk}v_lv_k\langle\Psi_0|P_l(t)P_k|\Psi_0\rangle$
with $P_l(t)=e^{i\hat{H_0}t}P_le^{-i\hat{H}_0t}$
using the quantum circuit for $\langle\Psi_0|P_l(t)P_k|\Psi_0\rangle$
shown in Fig. 1a repeatedly for different time $t$, and then performed a Fourier transform to obtain
$\chi(\omega)$. However, as the number of Pauli terms in $\hat{H}_0$ scales as $O(n^4)$ for
molecules, implementing $e^{-i\hat{H}_0t}$ requires a large circuit depth,
which at least scales as $O(n^4)$\cite{whitfield2011simulation,seeley2012bravyi,hastings2015improving} or $O(n^2)$ with low-rank approximations\cite{motta2021low}
asymptotically, which makes the application of these algorithms to molecules on near-term quantum hardware very challenging.

\begin{figure}[h!]
\centering
	\includegraphics[width=0.8\textwidth]{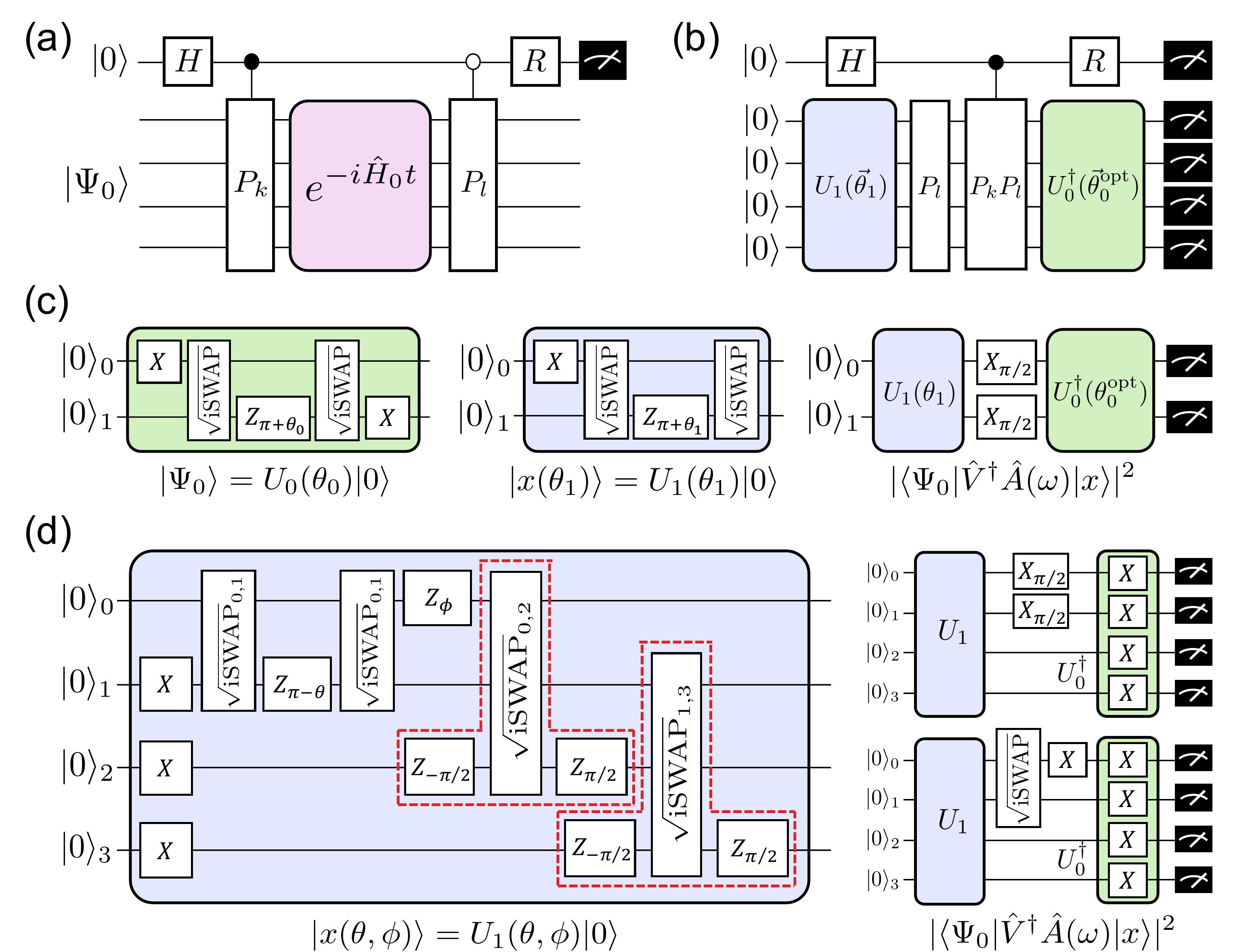}
	\caption{Quantum computation of linear response functions $\chi(\omega)$.
(a) Quantum circuit for computing the dynamical correlation function
$\langle\Psi_0|P_l(t)P_k|\Psi_0\rangle$ where $R$ is the Hadamard gate $H$ (or
$X_{\pi/2}$) for computing the real (or imaginary) part.
(b) Quantum circuit for computing
$\langle x|P_l|\Psi_0\rangle\langle\Psi_0|P_k|x\rangle$
with $|\Psi_0\rangle=U_0|0\rangle$ and $|x\rangle=U_1|0\rangle$.
(c) Two-qubit simulations of \ce{H2} and polyacenes: PQCs for $|\Psi_0\rangle$ and $|x\rangle$ as well as the quantum circuit for computing $|\langle\Psi_0|\hat{V}^\dagger\hat{A}(\omega)|x\rangle|^2$
(d) Four-qubit simulations of CO:
PQC for $|x\rangle$ and the two quantum circuits for
computing $|\langle\Psi_0|\hat{V}^\dagger\hat{A}(\omega)|x\rangle|^2$.
The red dashed box represents the entangling gate
generating a spin singlet pair $\frac{1}{\sqrt{2}}(|01\rangle+|10\rangle$)
between the two qubits.
}\label{fig:scheme}
\end{figure}

To overcome this difficulty, we adopt the frequency-domain formulation for response properties\cite{cai2020quantum,tong2021fast,chen2021variational},
which avoids the formidable sum over all excited states in Eq. \eqref{eq:SOS}
by introducing an auxiliary response state $|\Psi(\omega)\rangle$ satisfying
\begin{eqnarray}
\hat{A}(\omega)|\Psi(\omega)\rangle=\hat{V}|\Psi_0\rangle,
\quad\hat{A}(\omega)=\hat{H}_0-E_0-(\omega+i\gamma),\label{eq:rspw}
\end{eqnarray}
such that $\chi(\omega)$ becomes a simple expectation value
$\chi(\omega) = \langle \Psi(\omega)|\hat{A}^\dagger(\omega)|\Psi(\omega)\rangle$
similar to the energy $E_0=\langle\Psi_0|\hat{H}_0|\Psi_0\rangle$.
It has been shown that using the Harrow-Hassidim-Lloyd (HHL) algorithm\cite{harrow2009quantum}
or its improvements\cite{ambainis2010variable,childs2017quantum,subacsi2019quantum}
to solve the linear response equation \eqref{eq:rspw} can offer an exponential speedup\cite{cai2020quantum} over classical FCI-based response algorithms\cite{helgaker2012recent},
provided the ground state $|\Psi_0\rangle$ has been prepared on quantum computers.
However, realizing these quantum algorithms faces the same difficulty
as implementing the time evolution on NISQ devices.
Following the variational hybrid quantum-classical algorithm
for linear system of equations\cite{xu2019variational,bravo2019variational},
we present a pragmatic algorithmic primitive for solving Eq. \eqref{eq:rspw},
which can also be extended to high-order response equations for nonlinear response properties.
Suppose the ground state has been prepared by $|\Psi_0\rangle=U_0(\vec{\theta}_0^\mathrm{opt})|0\rangle$,
where $U_0(\vec{\theta}_0)$ is a parameterized quantum circuit (PQC) with a set of free angles $\vec{\theta}_0$
whose optimal values can be determined using the variational quantum eigensolver (VQE)\cite{peruzzo2014variational,mcclean2016theory}.
To solve Eq. \eqref{eq:rspw}, one can design a PQC $U_1(\vec{\theta}_1)$ for the normalized response state at a given frequency $|x\rangle\equiv|\Psi(\omega)\rangle/\||\Psi(\omega)\rangle\|=U_1(\vec{\theta}_1)|0\rangle$,
and then find the optimal parameters $\vec{\theta}_1^{\mathrm{opt}}$
by minimizing a cost function constructed using the Cauchy-Schwarz inequality
\begin{eqnarray}
L(\vec{\theta}_1)=\langle\Psi_0|\hat{V}^\dagger\hat{V}|\Psi_0\rangle
\langle x|\hat{A}^\dagger(\omega)\hat{A}(\omega)|x\rangle
-|\langle\Psi_0|\hat{V}^\dagger\hat{A}(\omega)|x\rangle|^2,\label{eq:costfun}
\end{eqnarray}
which obeys $L(\vec{\theta}_1)\ge 0$ . Its minimum is uniquely achieved
at $|x\rangle \propto \hat{A}^{-1}(\omega)\hat{V}|\Psi_0\rangle$ for the nonsingular
operator $\hat{A}(\omega)$. Note that the choice of cost functions for solving Eq. \eqref{eq:rspw} is not unique\cite{xu2019variational,bravo2019variational}, and our choice has the
advantage that the squared cross term $|\langle\Psi_0|\hat{V}^\dagger\hat{A}(\omega)|x\rangle|^2$
can be computed with less controlled operations (vide post), which are more friendly for NISQ devices. With $\vec{\theta}_1^{\mathrm{opt}}$ obtained at a given frequency,
$\chi(\omega)$ can be computed from (see Supporting Information\cite{SM})
\begin{eqnarray}
\chi(\omega)=
\langle x|\hat{A}^\dagger(\omega)|x\rangle \frac{\langle\Psi_0|\hat{V}^\dagger\hat{V}|\Psi_0\rangle}
{\langle x|\hat{A}^\dagger(\omega)\hat{A}(\omega)|x\rangle}.\label{eq:chiFinal}
\end{eqnarray}

The expectation values in Eqs. \eqref{eq:costfun}
and \eqref{eq:chiFinal}, i.e., $\langle\Psi_0|\hat{V}^\dagger\hat{V}|\Psi_0\rangle$,
$\langle x|\hat{A}^\dagger(\omega)\hat{A}(\omega)|x\rangle$,
and $\langle x|\hat{A}^\dagger(\omega)|x\rangle$, can be simply
computed by measurements after preparing $|\Psi_0\rangle$ or $|x\rangle$
on quantum computers. The computation of $|\langle\Psi_0|\hat{V}^\dagger\hat{A}(\omega)|x\rangle|^2$
in Eq. \eqref{eq:costfun} is more involved.
Using the expansion $\hat{V}^\dagger\hat{A}(\omega)=\sum_k c_k(\omega) P_k$
derivable from the expansions for $\hat{V}$ and $\hat{H}_0$,
$|\langle\Psi_0|\hat{V}^\dagger\hat{A}(\omega)|x\rangle|^2$
becomes $\sum_{lk} \bar{c}_l(\omega) c_k(\omega)
\langle x|P_l|\Psi_0\rangle\langle\Psi_0| P_k|x\rangle$.
We introduce a quantum circuit for computing
$\langle x|P_l|\Psi_0\rangle\langle\Psi_0|P_k|x\rangle$
with an ancilla qubit (Fig. 1b), whose depth is dominated by
the sum of depths for $U_0(\vec{\theta}_0^{\mathrm{opt}})$ and $U_1(\vec{\theta}_1)$.
Unlike previous works using the Hadamard test\cite{xu2019variational,bravo2019variational},
this scheme does not require any controlled operation on $U_0(\vec{\theta}_0^{\mathrm{opt}})$ nor $U_1(\vec{\theta}_1)$. Moreover, for certain problems with symmetries, the computation of $|\langle\Psi_0|\hat{V}^\dagger\hat{A}(\omega)|x\rangle|^2$
can be further simplified to quantum circuits without the ancilla (see Figs. 1c and 1d
as well as Supporting Information\cite{SM} for details).
With low-depth ans\"{a}tze\cite{kandala2017hardware} for $U_0(\vec{\theta}_0)$ and $U_1(\vec{\theta}_1)$, the present algorithm is more feasible on near-term quantum hardware
than existing quantum algorithms for response properties\cite{somma2002simulating,chiesa2019quantum,francis2020quantum,sun2021quantum}.

The present algorithm adds two new contributions
to the arsenal of quantum computational chemistry\cite{cao2019quantum,mcardle2020quantum,bauer2020quantum,motta2021emerging}.
First, the unique feature of VQR is that it enables the computation of linear response properties directly
in a frequency region of interest. The calculations for different frequencies are
completely independent and hence can be carried out in parallel.
Second, as a method for simulating excitation spectra,
VQR is more advantageous than quantum algorithms solely for excited states
\cite{mcclean2017hybrid,colless2018computation,higgott2019variational,nakanishi2019subspace,parrish2019quantum,ollitrault2020quantum}
in several aspects. It not only gives peak positions, but also provides
relative transition strengths simultaneously. Besides,
in the frequency region with a large density of states,
performing the slowly convergent sum in Eq. \eqref{eq:SOS}
by computing excited states is cumbersome,
and the implicit sum using VQR is more elegant.

{\it Hardware implementation and error mitigation.}
With the VQR algorithm, we are able to perform the first quantum simulation of
linear response properties of molecules on a superconducting processor,
including dynamic polarizabilities of the hydrogen molecule
(\ce{H2}), ultraviolet-visible (UV-Vis) absorption spectra of polyacenes,
and X-ray absorption spectra of carbon monoxide (CO).
In these cases, $\hat{V}$ is a dipole operator ($\hat{x}$, $\hat{y}$, or $\hat{z}$),
such that $\chi(\omega)$ represents the resonant
contribution to dipole polarizability, whose imaginary part is
related with the linear absorption of radiation $\sigma_{\mathrm{abs}}(\omega)$
by a randomly oriented molecular sample in the electric-dipole approximation\cite{norman2018principles},
viz., $\sigma_{\mathrm{abs}}(\omega)=\frac{4\pi\omega}{c}
\mathrm{Im}\bar{\chi}(\omega)$ with
$\bar{\chi}=\frac{1}{3}(\chi_{xx}+\chi_{yy}+\chi_{zz})$
and $c$ being the speed of light.
The lineshape of $\sigma_{\mathrm{abs}}(\omega)$ can be understood by noting that
$\mathrm{Im}\chi(\omega)=
\gamma\frac{\langle\Psi_0|\hat{V}^\dagger\hat{V}|\Psi_0\rangle}
{\langle x|\hat{A}^\dagger(\omega)\hat{A}(\omega)|x\rangle}
=
\sum_m |\langle\Psi_m|\hat{V}|\Psi_0\rangle|^2 \frac{\gamma}{(\omega_{m0}-\omega)^2+\gamma^2}$
is a weighted superposition of Lorentzians with a common full width at half maximum (FWHM) $2\gamma$.

The superconducting quantum processor\cite{Song2019Generation} used in our simulations
consists of 20 frequency-tunable transmon qubits connected via a central resonator,
which was employed before for realizing quantum generative adversarial networks\cite{huang2021QGAN}
and characterizing multiparticle entangled states\cite{xu2022metrological}.
A special feature of this device is that the $\sqrt{\textrm{iSWAP}}_{i,j}$ gate
can be realized between any pair of qubits.
The fidelity characterized by quantum process tomography\cite{Chuang1997,qguo2018}
is 0.9806 for the $\sqrt{\textrm{iSWAP}}$ gate in the simulations of \ce{H2} and polyacenes
with circuits shown in Fig. 1c, while the average fidelity for the three $\sqrt{\textrm{iSWAP}}$ gates in the simulations of CO with circuits shown in Fig. 1d is slightly lower (ca. 0.9651). More detailed information about the device and experimental setup can be found in Supporting Information\cite{SM}.

Since we mainly focus on the feasibility of VQR on NISQ devices in this work,
we employ PQCs capable of encoding the exact ground or response state as variational ans\"{a}tze, and
perform numerical optimizations by scanning the entire parameter space
(instead of using a classical optimizer).
To mitigate the impact of noises, we employ a simple EM strategy
based on symmetry projection for expectation values in VQE/VQR,
which is similar to the EM strategy by symmetry verification\cite{bonet2018low,sagastizabal2019experimental}.
Specifically, to improve the estimate of $\langle\Psi|\hat{O}|\Psi\rangle$ for
a quantum state $|\Psi\rangle$ with certain symmetry and
an operator $\hat{O}$ (e.g., $\hat{H}_0$,
$\hat{A}(\omega)$, $\hat{A}^\dagger(\omega)\hat{A}(\omega)$, and $\hat{V}^\dagger\hat{V}$)
commuting with the associated symmetry projector $\mathcal{P}$
(viz., $\mathcal{P}|\Psi\rangle=|\Psi\rangle$ and $[\hat{O},\mathcal{P}]=0$),
instead of using the raw result $\langle\Psi|\hat{O}|\Psi\rangle^{\mathrm{raw}}\equiv\mathrm{tr}(\rho_{\mathrm{exp}}\hat{O})$,
we can use $\langle\Psi|\hat{O}|\Psi\rangle^{\mathrm{EM}}\equiv\mathrm{tr}(\rho_{\mathrm{proj}}\hat{O})
=\mathrm{tr}(\mathcal{P}\rho_{\mathrm{exp}}\mathcal{P}\hat{O})/\mathrm{tr}(\rho_{\mathrm{exp}}\mathcal{P})
=\mathrm{tr}(\rho_{\mathrm{exp}}\hat{O}\mathcal{P})/\mathrm{tr}(\rho_{\mathrm{exp}}
\mathcal{P})$, where $\mathrm{tr}(\rho_{\mathrm{exp}}\hat{O}\mathcal{P})$ and $\mathrm{tr}(\rho_{\mathrm{exp}}\mathcal{P})$ need to be measured on quantum computers
and
$\rho_{\mathrm{proj}}=\mathcal{P}\rho_{\mathrm{exp}}\mathcal{P}/
\mathrm{tr}(\rho_{\mathrm{exp}}\mathcal{P})$
is the symmetry-projected density matrix with unphysical 
components outside the correct subspace
due to noises removed via symmetry projection. The specific form of $\mathcal{P}$ for each molecule can be found in Supporting Information\cite{SM}.
The use of EM is found essential for the accuracy of the simulated spectra,
since the demoninator $\langle x|\hat{A}^\dagger(\omega)\hat{A}(\omega)|x\rangle$
in Eq. \eqref{eq:chiFinal} is crucial for determining peak positions and heights.

\begin{figure}[h!]
\centering
	\includegraphics[width=0.8\textwidth]{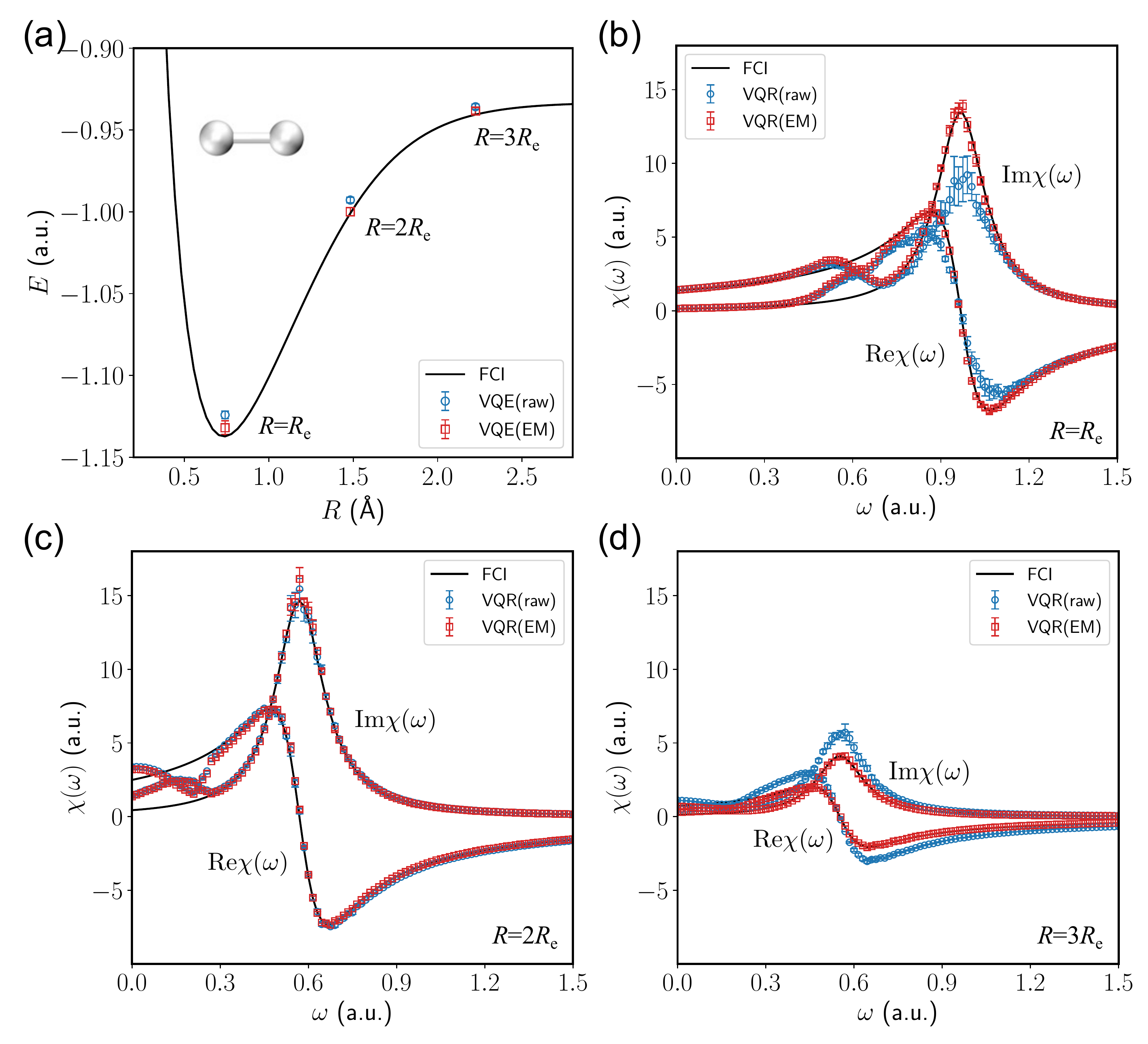}
	\caption{Ground-state energies and dynamic polarizabilities of \ce{H2}
at three representative bond distances from two-qubit simulations.
(a) Ground-state energies computed using VQE without (blue) and with EM (red).
(b),(c),(d) Real and imaginary parts of dipole polarizability
$\chi(\omega)$ for $\hat{V}=\hat{z}$ and $\gamma=0.1$ a.u.
computed using VQR without (blue) and with EM (red) at different bond distances.}\label{fig:h2}
\end{figure}

{\it Results.}
Due to its conceptual simplicity, \ce{H2} in a minimal basis set has become a test bed for new quantum algorithms\cite{lanyon2010towards,peruzzo2014variational,o2016scalable,kandala2017hardware,
hempel2018quantum,sagastizabal2019experimental}.
We compute its dipole polarizability for $\hat{V}=\hat{z}$ and $\gamma=0.1$ a.u. using VQR at three representative
bond distances, viz., $R=R_e$ (the equilibrium bond length\cite{huber1979}),
$2R_e$, and $3R_e$, which cover the weak, intermediate, and strong electron correlation regimes.
Figure 2 displays the ground-state energies and dynamic polarizabilities obtained by VQE and VQR, respectively,
using the quantum circuits shown in Fig. 1c.
In both VQE and VQR, EM with a spatial symmetry projector leads to a better agreement with the corresponding FCI values
than the raw results. This is particular the case for the peaks of $\mathrm{Im}\chi(\omega)$ predicted by VQR
for the transition from the ground state $|{}^1\Psi_g\rangle$
to the first singlet excited states $|{}^1\Psi_u\rangle$.
Meanwhile, small deviations from the FCI curves are observed for $\chi(\omega)$ at $\omega$
around 0.6, 0.1, and 0.0 a.u. for $R=R_e$, $2R_e$, and $3R_e$, respectively.
These frequencies turn out to be exactly the position of the first triplet excited state $|{}^3\Psi_u\rangle$ at each geometry (see Supporting Information\cite{SM}). Around these regions, the two terms in $L(\theta_1)$ are very close to each other
and the difference is on the order of $\gamma^2$.
Consequently, the computation of $L(\theta_1)$ is prone to noises on each term, such that the optimized angle $\theta_1^\mathrm{opt}$ becomes less accurate. Improving the two-qubit gate fidelity will alleviate this problem (see Supporting Information\cite{SM}),
and further applying a spin symmetry projector to filter out triplet states
will completely remove the small spurious speaks in $\mathrm{Im}\chi(\omega)$.
Comparison with $\chi(\omega)$ computed using the Hartree-Fock method and
DFT with some popular exchange-correlation functionals (see Supporting Information\cite{SM})
reveals the limitation of these approximate methods in the strong correlation
regime, for which VQR may become a better computational tool in future.

\begin{figure}[h!]
\centering
	\includegraphics[width=0.8\textwidth]{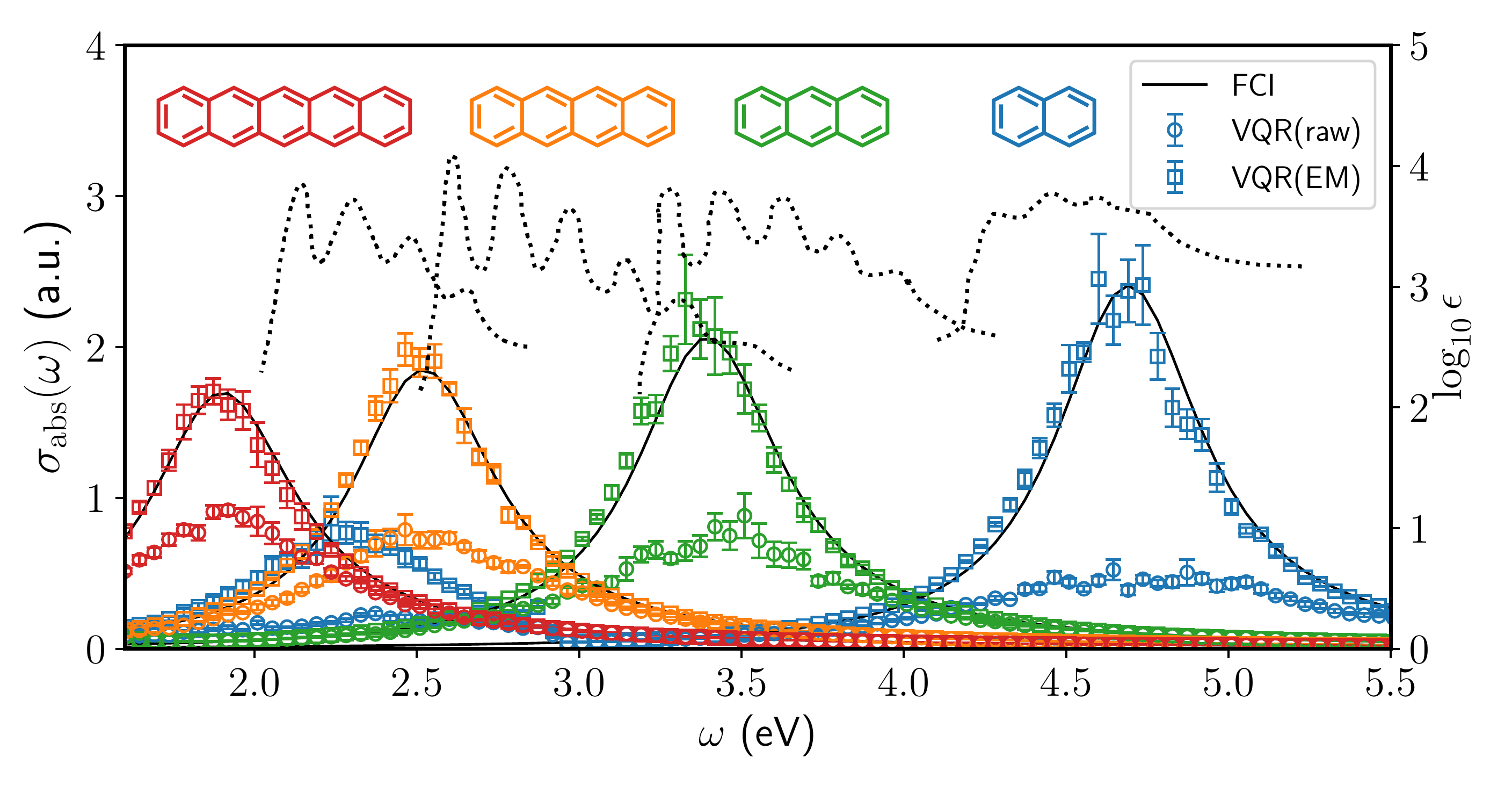}
	\caption{
UV-Vis absorption spectra $\sigma_{\mathrm{abs}}(\omega)$ ($\gamma=0.01$ a.u.) from
two-qubit simulations using VQR for polyacenes.
Experimental absorption spectra\cite{clar1964polycyclic} (dotted lines) are shown for comparison.
All the computed spectra have been shifted by -1.9 eV estimated using perturbation theory
for dynamic electron correlation (see Supporting Information\cite{SM}).
}\label{fig:polyacenes}
\end{figure}

Next, we examine the potential of VQR in solving important realistic problems.
We apply VQR to simulate the first absorption band
(the so-called $^1L_a$ or $p$ band\cite{clar1964polycyclic})
of polyacenes due to the transition from the highest occupied molecular orbital (HOMO)
to the lowest unoccupied molecular orbital (LUMO), which is the most relevant band for their applications as optoelectronic devices\cite{anthony2006functionalized}. As qubit resources are currently quite limited, simulating the entire molecule
remains impossible for polyacenes. Thus, we combine VQR with the complete active space (CAS) model\cite{szalay2012multiconfiguration},
in which only a selected number of electrons and orbitals are treated at the FCI level.
Figure 3 shows the computed UV-Vis absorption spectra $\sigma_{\mathrm{abs}}(\omega)$
for polyacenes (including naphthalene, anthracene, tetracene, and pentacene)
using a minimal active space composed of HOMO and LUMO with two active electrons,
denoted by CAS(2e,2o). The same quantum circuits (Fig. 1c) as for \ce{H2} are used in VQE and VQR
for the reduced active-space problems, but a smaller $\gamma$ (0.01 a.u.) is used
for a better resolution in the UV-Vis region.
Compared with the VQR(raw) results for \ce{H2},
the VQR(raw) results here show less good agreement with
the FCI references, because the computation of the denominator $\langle x|\hat{A}^\dagger(\omega)\hat{A}(\omega)|x\rangle$
in Eq. \eqref{eq:chiFinal} is more sensitive to noises
for smaller $\gamma$.
Even in this case, the VQR(EM) spectra agree well with the FCI spectra,
except for the appearance of a small spurious peak around 2.3 eV for naphthalene
due to the same problem found for \ce{H2}.
Compared with the experimental spectra of polyacenes\cite{clar1964polycyclic},
our simulations successfully reproduce the remarkable red shift
as the number of rings increases. To have a better quantitative agreement,
dynamic electron correlation and vibronic couplings need to be
taken into account in future. Although not illustrated in this work, we mention that the VQR algorithm is also applicable to simulating emission spectra,
provided the initial excited state is prepared on quantum computers,
e.g., using the variational quantum deflation method\cite{higgott2019variational}.

\begin{figure}[h!]
\centering
	\includegraphics[width=0.8\textwidth]{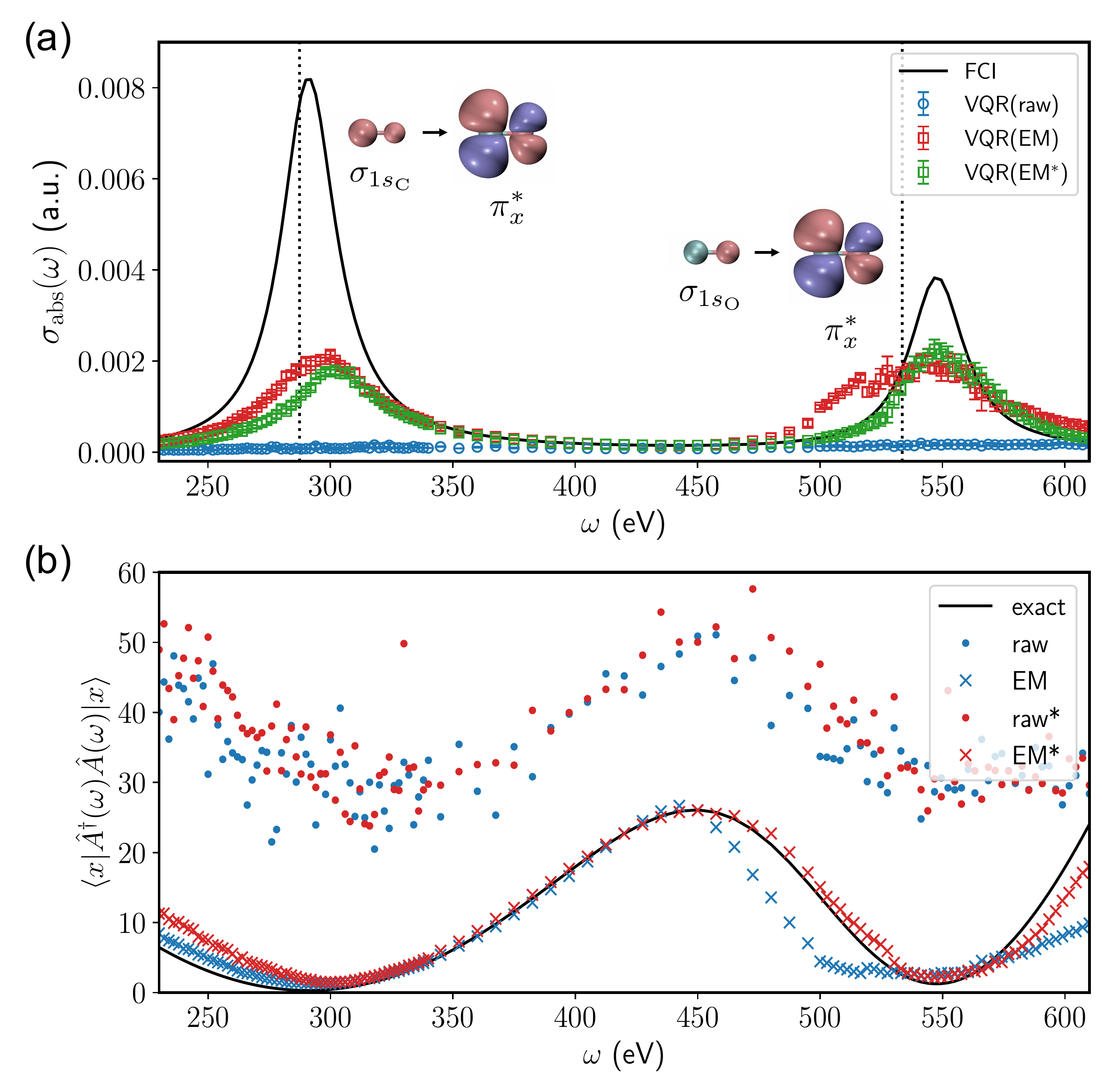}
	\caption{X-ray absorption spectra of CO from four-qubit simulations.
(a) Simulated spectra $\sigma_{\mathrm{abs}}(\omega)$ ($\hat{V}=\hat{x}$ and $\gamma=0.5$ a.u.)
using VQR. VQR(EM$^*$) represents the spectrum measured at the theoretical
angles $(\theta^*,\phi^*)$ at each frequency.
Experimental core excitation energies\cite{ma1991high,puttner1999vibrationally} (vertical dotted lines)
are shown for comparison.
(b) Results for $\langle x|\hat{A}^\dagger(\omega)\hat{A}(\omega)|x\rangle$
using the experimentally optimized $(\theta^{\mathrm{opt}},\phi^{\mathrm{opt}})$
(labelled by raw and EM) and the theoretical
$(\theta^{*},\phi^{*})$ (labelled by raw$^*$ and EM$^*$) at each frequency,
compared against the exact curve (black line).
}\label{fig:CO}
\end{figure}

Finally, we apply VQR to simulate X-ray spectroscopies, which are widely utilized for
probing local molecular and electronic structures, but remain
challenging for theoretical prediction\cite{norman2018simulating,besley2021modeling}.
The carbon K-edge and oxygen K-edge absorption spectra of CO
are simulated using VQR with a CAS(4e,3o) model composed of
$\sigma_{1s_{\mathrm{O}}}$, $\sigma_{1s_{\mathrm{C}}}$, and $\pi^*_x$
orbitals. The ground state is well-approximated by the Hartree-Fock state,
$|\Psi_0\rangle=U_0|0\rangle$ with $U_0=X_0X_1X_2X_3$.
For $\hat{V}=\hat{x}$, a response state can be exactly parameterized by
$|x(\theta,\phi)\rangle=U_1(\theta,\phi)|0\rangle=\cos\frac{\theta}{2}\osinglet+\sin\frac{\theta}{2} e^{i\phi}\csinglet$ with two free angles $\theta$ and $\phi$,
where $U_1(\theta,\phi)$ is the PQC in Fig. 1d and
$\osinglet$ ($\csinglet$) represents the singlet excited state due to the
transition from the core orbital $\sigma_{1s_{\mathrm{O}}}$ ($\sigma_{1s_{\mathrm{C}}}$) to
the unoccupied orbital $\pi^*_x$.
Figure 4a displays the simulated spectra $\sigma_{\mathrm{abs}}(\omega)$ using
VQR. It is clear that the improvement by EM is essential.
While the VQR(raw) spectrum is almost vanishing,
the VQR(EM) spectrum has two obvious absorption peaks.
Compared with the previous two-qubit simulations,
the deviation of the VQR(EM) spectrum from the FCI reference is larger for peak heights.
To better understand this discrepancy, we also measure the
spectra using the theoretical angles $(\theta^*,\phi^*)$ for $L(\theta,\phi)$ at each frequency (denoted by VQR(EM$^*$) in Fig. 4a), and compare the corresponding values of
$\langle x|\hat{A}^\dagger(\omega)\hat{A}(\omega)|x\rangle$ in Eq. \eqref{eq:chiFinal}
obtained using the experimentally optimized $(\theta^{\mathrm{opt}},\phi^{\mathrm{opt}})$
and the theoretical angles (see Fig. 4b).
This comparison reveals that the deviation in peak heights is not dominated by
errors in the optimization of $L(\theta,\phi)$, but is mainly due to errors
in computing $\langle x|\hat{A}^\dagger(\omega)\hat{A}(\omega)|x\rangle$
arising from experimental imperfections such as the lowering of two-qubit gate fidelities
in realizing the quantum circuit for $|x(\theta,\phi)\rangle$ (Fig. 1d).
Therefore, a further reduction of the deviation can be
anticipated using better quantum hardware with higher two-qubit gate
fidelities or more sophisticated EM strategies developed for VQE\cite{endo2021hybrid} in computing $\langle x|\hat{A}^\dagger(\omega)\hat{A}(\omega)|x\rangle$.
Note that even though the peak heights in the present VQR(EM) calculations are not ideal,
the core excitation energies for $\osinglet$ and $\csinglet$
determined by the peak positions are still reasonably good.
The remaining discrepancy with the experimental core excitation energies
obtained from high-resolution photoabsorption spectra\cite{ma1991high,puttner1999vibrationally}
can be improved by employing a larger active space and a better basis set in future.

{\it Conclusion.} We introduced a quantum algorithm feasible on near-term quantum hardware
for computing molecular linear response properties. While the reported experimental results are not perfect due to the presence of noises, the validity and feasibility of the VQR algorithm
are clearly demonstrated. The results highlight that it is the combination of three key innovations, i.e., frequency-domain formulation for response properties, variational algorithms for linear response equation with low-depth quantum circuits, and error mitigation techniques, that makes the simulation of molecular linear response properties possible on NISQ devices for the first time. This work suggests that a large class of important dynamical response properties such as emission spectra, Green's functions, and nonlinear optical properties are accessible
on near-term quantum hardware using the VQR approach.
Moreover, the present approach is not limited to superconducting devices, but is also applicable
to other quantum computing platforms.
Many improvements from different aspects can be easily envisaged, such as
lowering the number of measurements, employing better EM techniques,
using quantum gradients in optimizations\cite{huang2021QGAN},
as well as improving quantum hardware. Therefore, we believe that
the present algorithm opens the door for applying quantum computing
for simulating static and dynamical response properties of
more complex molecules and condense-phase systems in future.

\begin{acknowledgement}
This work was supported by the National Natural Science Foundation of China (Grant Nos. 21973003, 21688102, T2121001, 11934018, 11904393, 92065114 and 12174207), the Strategic Priority Research Program of Chinese Academy of Sciences (Grant No. XDB28000000) and the Beijing Natural Science Foundation (Grant No. Z200009).
\end{acknowledgement}

\begin{suppinfo}
Additional theoretical derivations, algorithms for
the cross term, and experimental details
including device information, two-qubit simulations for
\ce{H2} and polyacenes, and four-qubit simulations for CO.
\end{suppinfo}

\bibliography{references}

\end{document}